# INVESTIGATE THE PERFORMANCE EVALUATION OF IPTV OVER WIMAX NETWORKS

Jamil M. Hamodi and Ravindra C. Thool

Information Technology Department,
Shri Guru Gobind Singhji Institute of Engineering and Technology (SGGS)
Swami Ramanand Teerth Marathwada University (SRTM)
Nanded, INDIA
`{jamil_hamodi, rcthool}@yahoo.com`

## ABSTRACT

*Deployment Video on Demand (VoD) over the next generation (WiMAX) has become one of the intense interest subjects in the research these days, and is expected to be the main revenue generators in the near future and the efficiency of video streaming over next generation 4G is the key to enabling this. We are considering video streaming for real time video was coded by different H.264.x codes (H.264/AVC, and SVC), and we consider an IP-Unicast to deliver this streaming video over WiMAX. Our approach investigates the performance evaluation of IPTV (VoD) over WiMAX networks. OPNET is used to investigate the performance of VoD over WiMAX. Results obtained from simulation indicate that SVC video codec is an appropriate video codec for video streaming over WiMAX.*

## 1. INTRODUCTION

Worldwide Interoperability for Microwave Access (WiMAX) technology is the only wireless system capable of offering high QoS at high data rates for IP networks. It supports data rates of 70Mbps over ranges of 50km with mobility support at vehicular speeds [1]. An attractive growth rates in WiMAX subscriber base and equipment revenues in market studies last years, 133 million subscribers will be supported at the end of 2012 [2]. There is an increasing trend to deploy WiMAX technology for offering different application, such as voice, data, video, and multimedia services. Each of these applications has different QoS requirements.

Internet Protocol Television (IPTV) has become popular as it promises to deliver the content to users whenever they want. IPTV is a set of multimedia services that are distributed throughout an IP network, where end of user receives video streams through a set-top-box (STB) connected to a broadband connection. IPTV is often combined with the services of VoD. VoD services contents are not live but pre-encoded contents available at any time from servers. These services must possess an adequate level of quality of service, security, interactivity, and reliability. From the perspective of the provider, IPTV includes the video acquisition, video processed and video secure distribution on the IP network infrastructure [3], [4].

WiMAX technology is one of the access technologies that enables transmission of IPTV services. Transmitting IPTV over WiMAX aims to make IPTV services available to users anywhere, anytime and on any device. The QoS for delivering IPTV services depends especially on network performance and bandwidth [5]. The generic network topology of the IPTV application over WiMAX is shown in Figure 1.

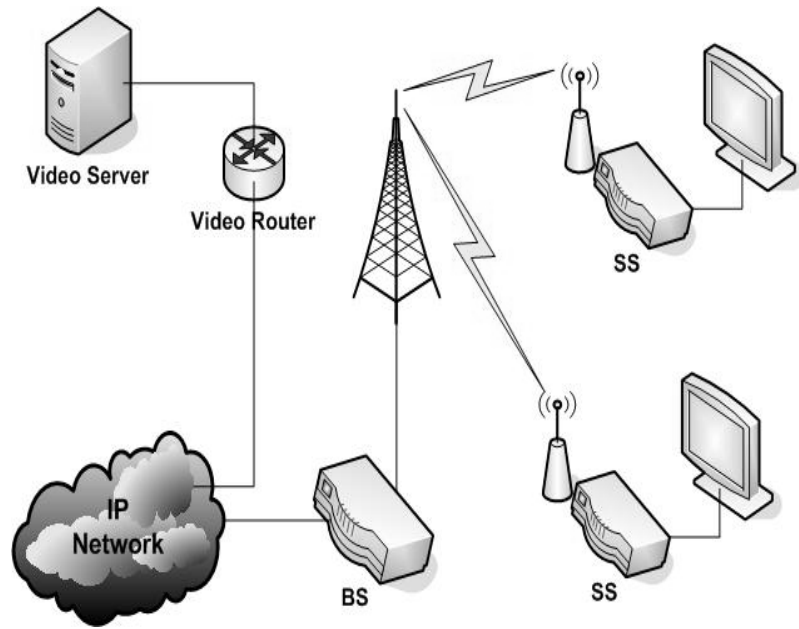

Figure 1. System Model for IPTV Application [5]

Scalable Video Coding (SVC) has achieved significant improvements in coding efficiency with an increased degree of supported scalability relative to the scalable profiles of prior video coding standards. Transmitting SVC encoded videos over WiMAX networks is an effective solution which solves many of the video transmission problems over these networks [6].

IEEE 802.16 Medium Access Control (MAC) specifies five types of QoS classes are [7]:

- ➢ Unsolicited grant services (UGS), which are designed to support constant bit rate services, such as T1/E1 and VoIP.

- ➢ Real-time polling services (rtPS), which are used to support real-time variable bit rate services such as MPEG video and VoIP.

- ➢ Non-real-time polling services (nrtPS), which are used to support non-real-time variable bit rate services such as FTP.

- ➢ Best-effort services, which enable to forward packets on a first-in-first basis using the capacity not used by other services.

- ➢ Extended real-time variable rate (ERT-VR) service, which is defined only in IEEE 802.16e, and is designed to support real time applications, that have variable data rates but require guaranteed data rate and delay.

Our proposed is to investigate the performance evaluation of IPTV (VoD) over WiMAX networks in order to investigate and analyze the behaviour and performance of the model. In sharp contrast to prior work in [2], this paper focuses on using different H.264.x codec's for modelling and simulation of the IPTV deployment over WiMAX. Secondly, our case study is focused on a more practical example pertaining on WiMAX networks. Finally, our approach measures the quality of video traffic for IPTV deployment over WiMAX network, using the Mean Opinion Score (MOS) metric, where a decreasing MOS value indicates poor video quality, as opposed to a higher value. The main objectives of this research are: To study the importance of IPTV (VoD), and to identify the factors affecting the VoD performance.

The rest of this paper is organized as follows: Section 2 presents related work that analyzes performance metrics of video over WiMAX. Section 3 presents a brief overview of WiMAX. An overview of the video content and its compression are given in Section 4. Section 5 describes performance issues and requirements that have to be defined for VoD service as a new service to be deployed. Section 6 describes the practical steps to be taken prior to simulation. Simulation results and analysis was obtained in Section 7. The study concludes in Section 8.

## 2. RELATED WORK

There are various related efforts have explored WiMAX in the context of real-time and stored video applications. Challenges for delivering IPTV over WiMAX were discussed in [7]. These include the challenges for QoS requirements. Also, they described the transmission of IPTV services over WiMAX technology, and the impact of different parameters in WiMAX network when deploying this service. An intelligent controller was designed based on fuzzy logic to analyze QoS requirements for delivering IPTV over WiMAX. Also, an intelligent controller based on fuzzy logic in [8] was used to analyze three parameters: jitter, losses and delays that affect the QoS for delivering IPTV services. The aim was to define a maximum value of link utilization among links of the network.

An OPNET Modeler was used to design, characterize, and compare the performance of video streaming to WiMAX and ADSL in [2], [9], [10]. The simulation results indicated that, ADSL exhibited behavior approached the ideal values for the performance metrics. While WiMAX demonstrated promising behavior within the bounds of the defined metrics. The work in [9] is extended work in [2] to include generation and integration of a streaming audio component, also enhanced the protocol stack to include the real time protocol (RTP) layer. Network topology was redesigned to incorporate WiMAX mobility. Also, they include characterization of WiMAX media access control (MAC) and physical (PHY) layer. Simulation scenarios used to observe the impact on the four performance metrics. Also, in [10] simulation used to compare the performance metrics between ADSL and WiMAX by varying the attributes of network objects such as traffic load and by customizing the physical characteristics to vary BLER, packet loss, delay, jitter, and throughput. Simulation results demonstrated considerable packet loss. ADSL exhibited considerably better performance than the WiMAX client stations.

An important performance issue was addressed when multimedia traffic is carried over WiMAX systems in [11]. The main focus is to show the effectiveness of QoS capabilities in delivering streaming multimedia such as IPTV and similar media content. The results provide a good indication on the applicability of WiMAX for multimedia applications. Measured and analyzed the performance of VoIP over WiMAX in terms of crucial parameters was presented in [12]. Different parameters such as jitter, MOS value, packet end-to-end delays and packets sent and received used to measure the performance of VoIP over WiMAX with different codes G.711, G.723, and G.729 in order to find the most appropriate voice codec for VoIP over WiMAX network. The simulation results showed that VoIP performed better under the G.711 codec as compared to the G.723 and G.729 codecs. The findings also showed that VoIP applications can perform better under the exponential traffic distribution.

A brief technical overview of SVC was presented in [13] when deployed on IPTV services. Based on this technical characterization, it is described how the different SVC features such as efficient methods for graceful degradation, bit rate adaptation and format adaptation, can be mapped to the application requirements of IPTV services. It is discussed how such mappings can lead to an improved content portability, management and distribution as well as an improved management of access network throughput resulting in better quality of service and experience for the users of IPTV services.

## 3. WIMAX OVERVIEW

In this Section, we give a brief overview of WiMAX technology. Worldwide Interoperability for Microwave Access (WiMAX) is currently one of the hottest technologies in wireless. The Institute of Electrical and Electronics Engineers (IEEE) 802 committee has published a set of standards that define WiMAX. IEEE 802.16-2004d was published in 2004 for fixed applications; 802.16e is publicated in July 2005 for mobility. WiMAX is a standard-based wireless technology that provides high throughput broadband connections over long distance. WiMAX can be used for a number of applications, including "last mile" broadband connections, hotspots and high-speed connectivity for business customers. It provides wireless metropolitan area network (MAN) connectivity at speeds up to 70 Mbps and the WiMAX base station on the average can cover between 5 to 10 km.

WiMAX operates in the 10–66 GHz band with line of sight (LOS) communications using the single carrier (SC) air interface. The IEEE 802.16a standard outlined non line of sight (NLOS) communications in the 2 – 11 GHz band using one of three air interfaces: SC, Orthogonal Frequency Division Multiplex (OFDM), and Orthogonal Frequency Division Multiple Access (OFDMA). OFDM and OFDMA enable carriers to increase their bandwidth and data capacity. This increased efficiency is achieved by spacing subcarriers very closely together without interference because subcarriers are orthogonal to each other [14], [15]. Channel bandwidths range between 1.25 MHz and 20 MHz in the 2 – 11 GHz with OFDM, the number of subcarriers scales linearly with the channel bandwidth. Within a given channel bandwidth, subcarriers are allocated as: null subcarriers, data subcarriers, pilot subcarriers, and DC subcarriers. Subcarriers are then modulated using conventional digital modulation schemes with various inner code rates: Binary Phase Shift Keying (BPSK), Quadrature Phase Shift Keying (QPSK), Quadrature Amplitude Modulation (QAM) (16-QAM, 64-QAM, and optional 256-QAM). Consequently, WiMAX data rates between 1.5 to 75 Mbps are achievable.

## 4. VIDEO OVERVIEW AND COMPRESSION

This section gives a brief overview of the video content and compression technology. The video information available from video service providers is referred to as the video content. The content is structured as a sequence of video frames or images that are sent or streamed to the subscriber and displayed at a constant frame rate [2]. Based on the differential transmission of the streaming real-time video and their buffering requirements from the network and the client server, video content is characterized by several parameters including video format, pixel color depth, coding scheme, and frame interval rate.

A numerous amount of compression technologies was developed by the International Standards Organization (ISO) and the International Telecommunication Union (ITU-T). The main advantage of this code and compression technology is that it has reduced the required bandwidth for encoding content that is done by reducing the space within the frame content. The series of MPEG specifications that are in widespread use in virtually. All existing IPTV deployments have been developed by jointly between the International Telecommunication Union (ITU-T) and the International Organization for Standards/International Electrotechnical Commission (ISO/IEC) [16]. The introduction of ITU-T recommendation H.264 or as known as an advanced video codec (AVC) has become interest to deploy future IPTV and VoD, which is developed in 2005 [17]. They proposed that a gamma distribution provides the best fit for the frame size histograms of H.264 codec video streams. H.264 (AVC) cuts the bandwidth requirement for digital video delivery to half of the MPEG-2. It uses only between 1-2 Mbps [16], [17], [18].

The H.264/SVC video codec which is the name for the Annex G extension of the H.264/MPEG-4 AVC, has been found to offer improved visual quality compared to the preceding standards and proved to be suitable for dynamic bandwidth environments. Telecommunications companies focus their efforts to deploy IPTV-DVD quality video services over DSL and the Internet by using H.264 (SVC) instead of H.264/ AVC and MPEG-x.

## 5. VIDEO TRAFFIC CHARACTERISTICS AND REQUIREMENTS

This section discusses some issues that come up when have been needed to deploy video in any network. These issues are characterizing the nature of video traffic, QoE, QoS requirements, and hardware components or devices.

### 5.1 Quality of Experience (QoE)

Quality of Experience is more important than quality of service (QoS). QoE is measured by the user's perception of the service. QoE for voice and video has traditionally been measured by Mean Opinion Score (MOS). MOS is based on the perception of the quality of conversation and picture of a sample population [19]. The quality of real-time services for video has been benchmarked against the Mean Opinion Score (MOS) mandated by the International Telecommunication Union, and Telecommunication Standardization Sector (ITU-T) for circuit switched networks.

Two standard methods used to evaluate the video quality: Calculating the PSNR between the source and the received (possibly distorted) video sequence the first method was used to assess the performance of video transmission systems, which is calculated image-wise and very similar to the well-known SNR but correlating better with the human quality perception. Quality impression of human observer's scale which showing in Table 1 was developed by ITU-R to illustrate the meaning of quality measures for non-experts [20]. BT.500 describes a methodology to gain these quality indicators by subjective assessment series (by a group of humans) was the second method. Heuristic mapping of PSNR to MOS values in [21] which showing in Table 2 can be used to roughly estimate the human quality perception for videos with relatively low motion.

Table 1. ITU-R quality and impairment

| Scale | Quality   | Impairment    |
|-------|-----------|---------------|
| 5     | Excellent | Imperceptible |
| 4     | Good      | Perceptible   |
| 3     | Fair      | Slightly      |
| 2     | Poor      | Annoying      |
| 1     | Bad       | Very annoying |

Table 2. PSNR to MOS conversion

| PSNR [dB] | MOS             |
|-----------|-----------------|
| > 37      | > 5 (Excellent) |
| 31 – 37   | 4 (Good)        |
| 25 – 31   | 3 (Fair)        |
| 20 -25    | 2 (Poor)        |
| < 20      | < 1 (Bad)       |

### 5.2 Quality of Service (QoS)

Quality of Service (QoS) is very important for deploying IPTV and VoD as it is a real-time service. However, QoS for deploying IPTV will be affected by packet loss, packet end-to-end delay, throughput, and jitter.

**Packet End-to-end delay**: Small amount of delay does not directly affect the Quality of Experience (QoE) of IPTV. While the delay large than 1 second may result a much worse QoS toward end-user experience. The delay for one way must be less than 200ms. On the other hand, the end-to-end delay more than 400ms was considered to be unacceptable [22].

**Packet loss**: Average number of packets lost compared to send a packet per second. Packet lost in the video processing layer is used for comparison purpose. There are some standard packet loss ratios given by ITU-T for classifying IPTV services. The excellent service quality of packet loss ratio is than $10^{-5}$ and the poor service quality of packet loss ratio is between $2*10^{-4}$ and 0.01 [22], a packet loss ratio above 1% is unacceptable.

**Jitter**: Calculated as the signed maximum difference in one way delay of the packets over a particular time interval [23]. Generally, jitter is defined as the absolute value of delay difference between selected packets. The jitter delay for one way must be less than 60ms on average and less than 10 ms in ideal [2].

**Throughput**: the rate at which a computer or network sends or receives data. It therefore is a good measure of the channel capacity of a communication link and connections to the internet are usually rated in terms of how many bits they pass per second (bit/s). The minimum end-to-end transmission rate acceptable for video is between 10 kbps and 5 Mbps [2].

### 5.3 Video equipments

The main equipments needed to deploy VoD or IPTV services are set-top boxes and a head-end server. The head-end server or known as VoD server is the source for all video content. The main functionality of set-top box (STB) is to unscramble the signal and present it on the TV [24].

## 6. SIMULATION SETUP

This section describes the simulation model adopted for analyzing the effect of Video on Demand (VoD) over the WiMAX networks .The simulation was performed to evaluate the performance of VoD over the WiMAX networks. Our simulation approach used the popular MIL3 OPNET Modeler simulation package1, Release14.0.A [25]. We used the OPNET Modeler to facilitate the utilization of in-built models of commercially available network elements, with reasonably accurate emulation of various real life network topologies.

The network topology of our test bed network is given in Figure 2. Our model consists of a circular placement of nodes in a hexagon with one WiMAX Base Station, and five Subscriber Stations (SS) which were 1km apart from the Base Station (BS). These nodes are fixed nodes as there was no mobility configured. The BS connected to the IP backbone via a DS3 WAN link, our video server connected to the server backbone via ppp_sonet_oct1 link.

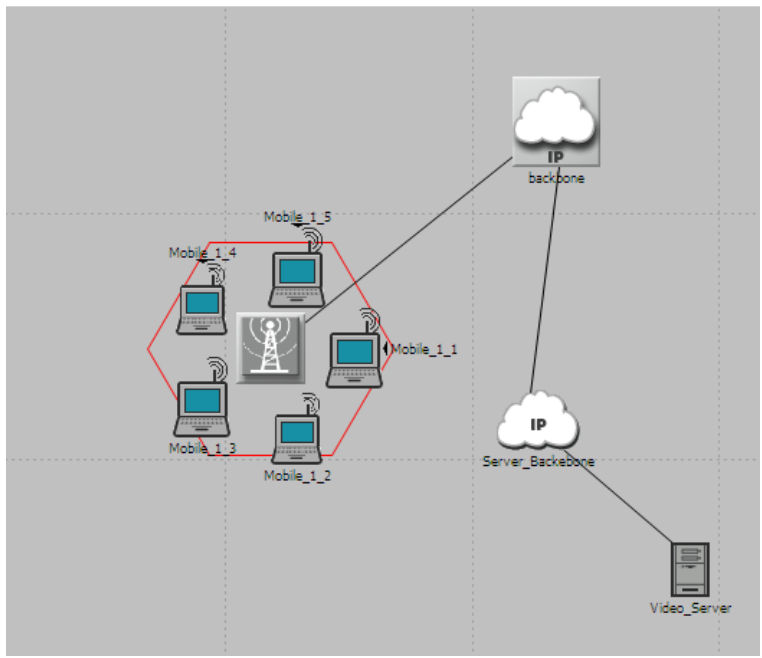

Figure 2. OPNET Model of WiMAX network

Different scenarios have been designed in our simulation with different video codes to identify the most appropriate video codec, and to measure the performance of some parameters such as throughput, jitter, packet loss and packet end-to-end delay as well as to meet the goal to analyze the quality of service of IPTV (VoD) over WiMAX networks.

Video streaming over wireless networks is challenging due to the high bandwidth required and the delay sensitive nature of video than most other types of application. As a result, a video traces of Tokyo Olympics with different video codes were used in our simulation. These video codes are H.264/AVC based scheme, and SVC based scheme used for comparison. These traffics were obtained from Arizona State University [26], [27], with a 532x288 from resolution, Group of Picture (GOP) size is selected as 16 for this video for all codes, and encoded with 30 frames per seconds (fps). Table 3 shows the mean and peak rates for these video codes of the Tokyo Olympics movie; all the video traces reflect only video frames.

Table 3. Video codec traces characteristics

| Parameters | H.264/AVC | SVC |
|---|---|---|
| Frame Compression Ratio | 21.7 | 18.01 |
| Min Frame Size (Bytes) | 17 | 22 |
| Max Frame Size (Bytes) | 62269 | 58150 |
| Mean Frame Size (Bytes) | 7004.52 | 8440.74 |
| Peak Frame Rate (Mbps) | 14.92 | 13.9 |
| Mean Frame Rates (Mbps) | 1.68 | 2.02 |
| Mean Frame PSNR (dB) | 46.49 | 47.89 |

## 6.1 Video Application Configuration

To the best of our knowledge, the OPNET Modeler does not have built in features to support video streaming or its deployment. In this subsection, we elaborate on our approach for emulating video streaming traffics. The parameters of the video conferencing application in

International Journal of Computer Networks & Communications (IJCNC) Vol.5, No.1, January 2013

modeler are: The frame inter-arrival time and the frame size. The incoming frame inter-arrival rate was configured to reflect the content encoding rate of 30 fps, and the outgoing stream inter-arrival time remains at none to create a unidirectional stream. The video traces were scripted into video conferencing frame size as shown in Figure 3.

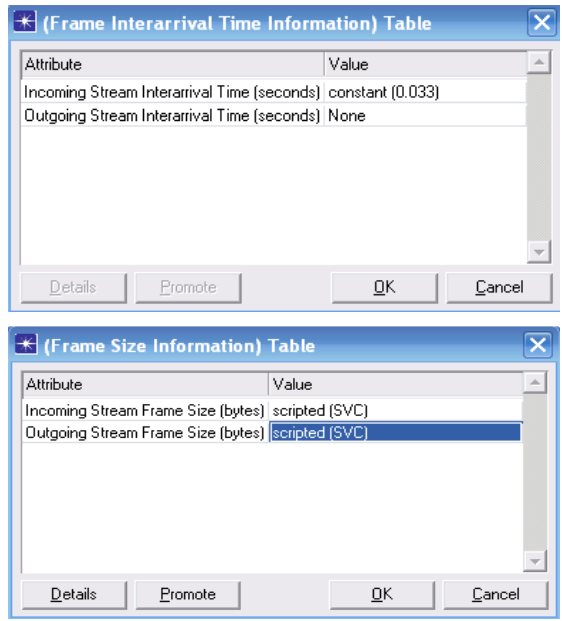

Figure 3. Application Configuration of video traffic

Figure 4 shows the modeler profile which configured in OPNET to reflect the video streaming defined applications (H.264/AVC, and SVC) codes traces of the Tokyo Olympics. The operation mode for this profile was configured to be simultaneous, with a starting time of 70 seconds. The video clients are subsequently configured with this profile, and the VoD server is configured to support these appropriate application services.

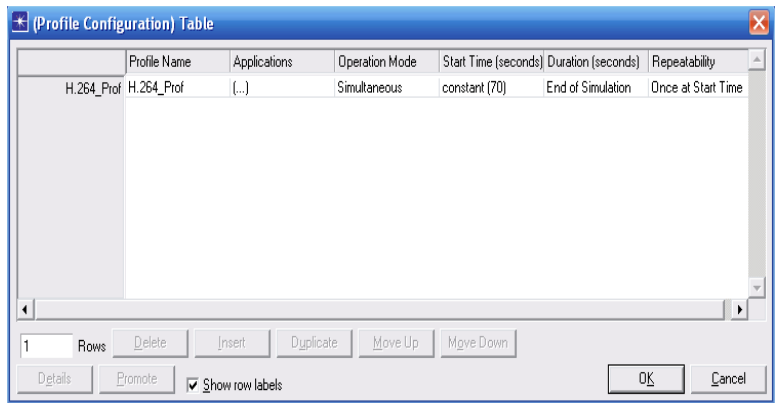

Figure 4. H.264/SVC video streaming traffic profile configuration

## 6.2 WiMAX Configuration

Our WiMAX model was configured to support video traffic on only fixed station which means not supporting mobility. (rtPS) scheduling class was created for the downlink and uplink to support the real time video streaming. The scheduling was configured with 5 Mbps Maximum sustainable traffic rate, and 1 Mbps Minimum sustainable traffic rate as shown in Figure 5.

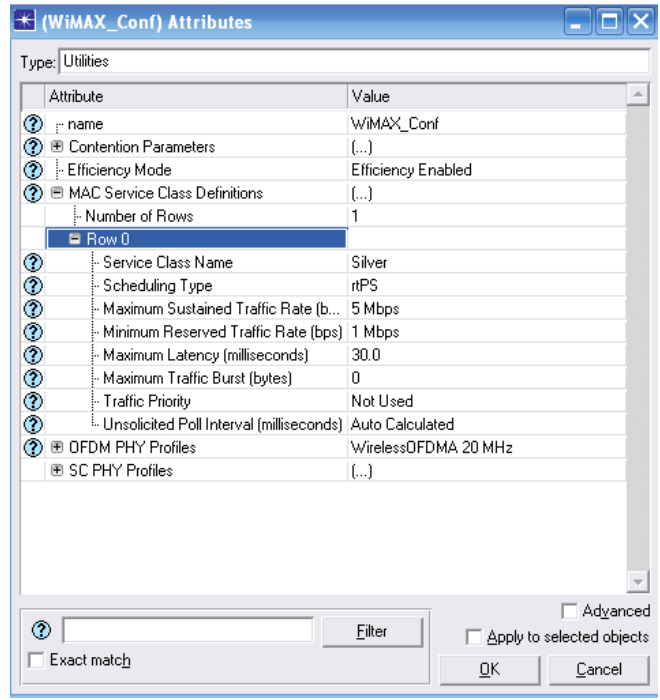

Figure 5. WiMAX classes Configuration

The Base Station and WiMAX Subscriber Stations were configured to map the uplink and down link service flows to a specific type of service (ToS) setting that was configured during the application node configuration. Moreover, each service flow uplink and downlink can be configured with the specific burst profile. For this study, the uplink channel was configured with 16-QAM which is different burst profile than of downlink 64-QAM. Figure 6, and Figure 7 show Base Station, and WiMAX Subscriber Stations configuration attribute.

Physical (PHY) layer access was configured to utilize OFDM over a 2.5 GHz base frequency using a 5 MHz channel bandwidth that provisions 512 subcarriers allocated. The base station transmit power was configured to 35.8 dB which equals approximately 3.8 watts with 15 dBi gain antenna. On the other hand, the client station transmit power was configured to use 33 dB equivalent to about 2 watts of transmit power over the 5MHz channel bandwidth, using 14 dBi gain antennas. Moreover, a fixed suburban pathloss model was employed with a conservative terrain model that accounted for mostly flat terrain with light tree densities.

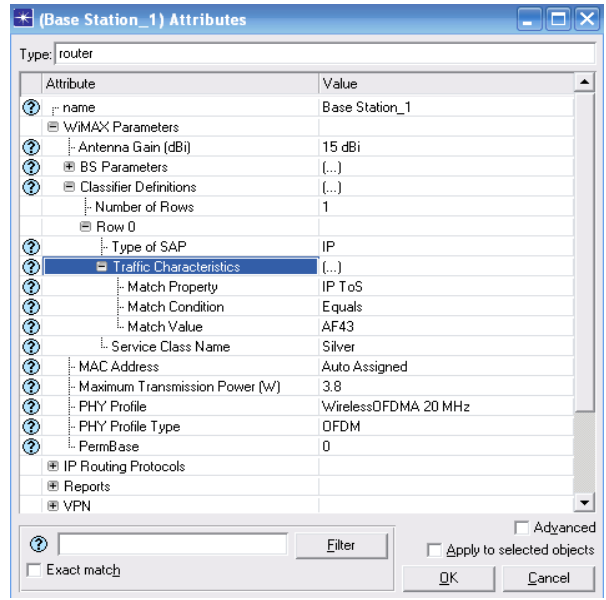

Figure 6. Base Station Parameters

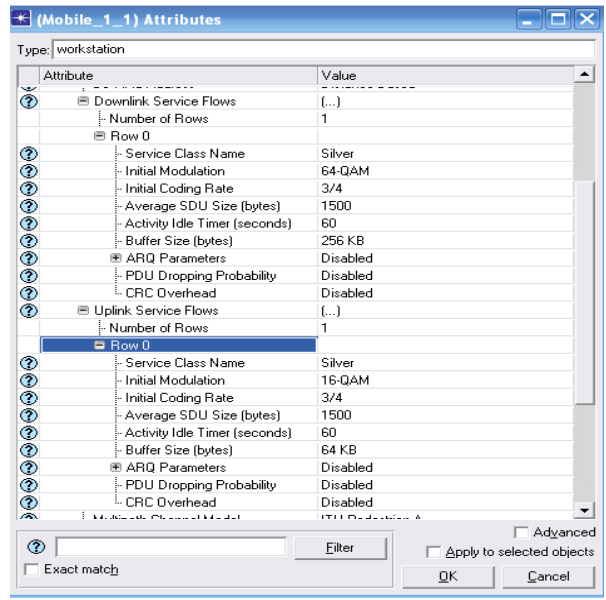

Figure 7. Subscriber Station modulation and Coding rate

## 7. SIMULATION RESULTS AND ANALYSIS

Snapshots of OPNET simulation results for deploying VoD services with different codes over WiMAX model was presented in this section. The goal of our simulation was to test the deployment, and to analysis the performance metrics IPTV (VoD) over WiMAX networks. The simulation results of our model are averaged across the 74 minute movie. Average values of end-to-end delay, jitter, and throughput are used for the analysis in all the Figures. The results obtained from these codes are similar. So, we obtained only the results of SVC codec.

International Journal of Computer Networks & Communications (IJCNC) Vol.5, No.1, January 2013

In Figure 8, we illustrate the corresponding streaming video end-to-end delay. This is expected as delay represents the average time of transit of each packet for more users and in it theory would be more as a lot more traffic is sent for multiple users. The average value for the end-to-end delay is 200ms, and in our case the delay is significantly less than that as shown by the curve to be 11ms for SVC.

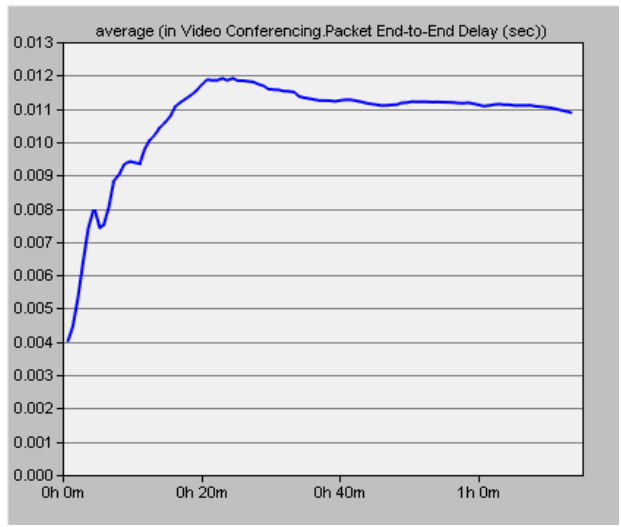

Figure 8. Averaged End-to-End packet delay

As we know, the ideal jitter is less than 10ms, which is the Packet Delay Variation for the video traffic. In our case the average jitter value for multiple users is shown in Figure 9 and is about 60 µs. We can observe that the jitter value for our case is significantly less than the ideal value of 10ms which is considered as a robust statistic

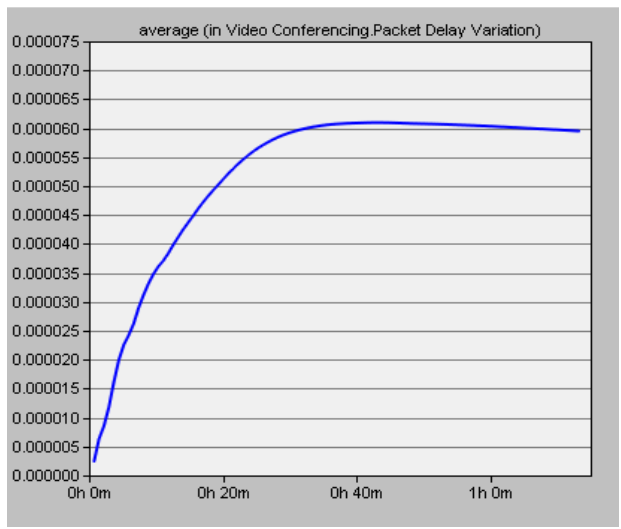

Figure 9. jitter delay

The throughput for our case in the range required of 10 kbps – 5 Mbps. The average throughput for our case is about 1.5 Mbps as shown in Figure 10. This indicates the throughput of our video codec is as expected.

International Journal of Computer Networks & Communications (IJCNC) Vol.5, No.1, January 2013

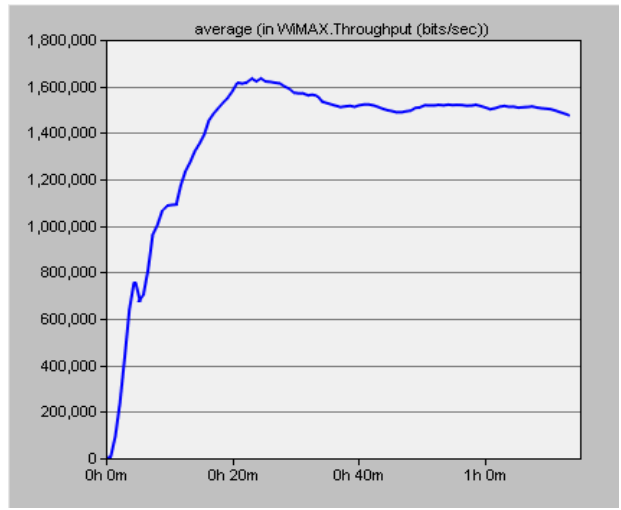

Figure 10. average throughput

Table 4. Modulation/coding rates

| Modulation | Coding | Information Bits/symbol/Hz | Required SNR (dB) |
|---|---|---|---|
| QPSK | ½ | 1 | 9.4 |
|  | ¾ | 1.5 | 11.2 |
| 16-QAM | ½ | 2 | 16.4 |
|  | ¾ | 3 | 18.2 |
| 64-QAM | ²/₃ | 4 | 22.7 |
|  | ¾ | 4.5 | 24.4 |

The dropped packet rates by the PHY layer for the WiMAX Subscriber Station are shown in Figure 11 (a), which exhibits a much higher loss rate. The downlink SNR for the Subscriber Station is shown in Figure 11 (b). The subscriber station exhibits a downlink SNR that is below the necessary minimum level for 64-QAM with ¾ coding which showing in Table 4. The low SNR for the subscriber station is a major contributor to the high packet loss rate.

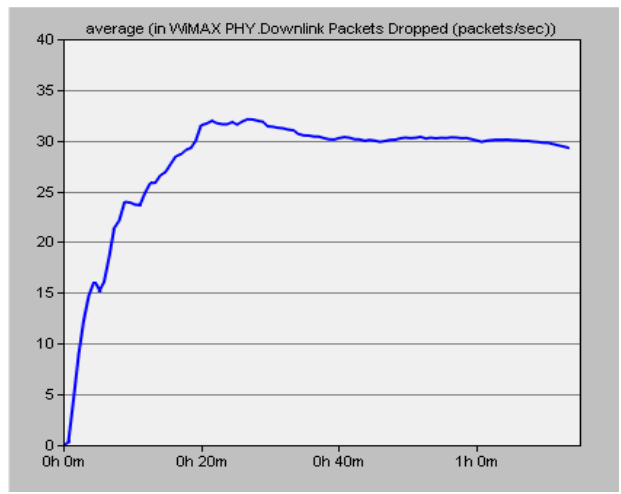

(a)

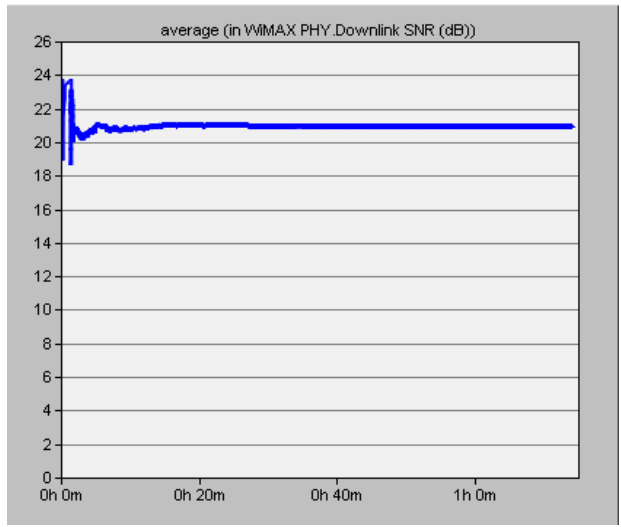

(b)

Figure 11. PHY layer; (a) lost packet and (b) SNR

Quality of Experience has been measured by MOS. Mean Opinion Score (MOS) is dependent on calculating the Peak Signal Noise Ratio (PSNR). We used the PSNR ratio from the State University [26] as in Table 3. From that result, we can see all these codes have a good MOS, and PSNR for SVC code is about 47.89, which means it has an excellent MOS.

The simulation results clearly indicate that SVC provides the best quality of video in terms of MOS value, throughput, end-to-end delays and jitters. Therefore, SVC the most appropriate video codec scheme for delivering IPTV services over WiMAX network. The main findings of this case are listed in Table 5.

Table 5. performance metrics SVC Video codec

| Parameters | SVC |
| --- | --- |
| Throughput | 1.25 Mbps |
| End-to-End Delay | 2.7 ms |
| Jitter Delay | 5.6 µs |
| PSNR (dB) [3] | 47.89 |

## 8. CONCLUSION

This study explores the technical details and performance analysis of IPTV over WiMAX broadband access technology. Its aim is to address the performance metrics of QoS for video streaming when deploying over WiMAX access technology. The OPNET Modeler is used to design and characterize the performance parameters of Tokyo Olympics video streaming with different codes of H.264.x to WiMAX video subscribers using QoS performance metrics. The simulation results indicate that, the H.264/SVC video codec has been found to offer improved visual quality and appropriate codes for delivering video compared to the preceding standards. Furthermore, the streaming video content has been modeled as unicast traffic while multicast video traffic may have yielded better performance. This work has limitations to certain assumptions like: Station transmit power, distance between base station and subscriber station, subscriber station was configured as fixed not support mobility, station antenna gain, carrier operating frequency and channel bandwidth.

International Journal of Computer Networks & Communications (IJCNC) Vol.5, No.1, January 2013## REFERENCES

[1] I. Uilecan, C. Zhou, and G. Atkin, "Framework for delivering IPTV services over WiMAX wireless networks," Proc. IEEE EIT 2007, Chicago, IL, May 2007, Vo. , NO. , pp. 470–475.

[2] W. Hrudey and Lj. Trajkovic, "Streaming video content over IEEE 802.16/WiMAX broadband access," OPNETWORK 2008, Washington, DC, Aug. 2008, Vo. , NO. , PP.

[3] Wikipedia. Available at: http://en.wikipedia.org/wiki/IPTV

[4] IPTV Focus Group. Available at: http://www.itu.int/ITUT/IPTV

[5] F. Retnasothie, M. Ozdemir, T. Yucek, H. Celebi, J. Zhang, and R. Muththaiah, "Wireless IPTV over WiMAX: challenges and applications," Proc. IEEE WAMICON 2006, Clearwater, FL, Dec. 2006, pp. 1–5

[6] K. S. Easwarakumar, and S. Parvathi, "Performance Evaluation of Multicast Video Streaming over WiMAX.", International Journal of Applied Information Systems, 2012, Vo.3, No. 4, pp.

[7] A. Shehu, A. Maraj, and R.M. Mitrushi, "Analysis of QoS requirements for delivering IPTV over WiMAX technology," International Conference on Software, Telecommunications and Computer Networks (SoftCOM), 2010, vol., no., pp. 380-385.

[8] A. Shehu, A. Maraj, and R.M. Mitrushi, "Studying o different parameters that affect QoS in IPTV systems," International Conference on Telecommunications and Information (WSEAS), 2010, vol., no., pp.

[9] W. Hrudey and Lj. Trajkovic, "Mobile Wimax MAC and PHY layer optimization for IPTV," Journal of Mathematical and Computer Modeling, Elsevier, Mar. 2011, vol. 53, pp. 2119–2135.

[10] R. Gill, T. Farah, and Lj. Trajkovic, "Comparison of WiMAX and ADSL performance when streaming audio and video content," OPNETWORK 2011, Washington, DC, Aug. 2011.

[11] D.J. Reid, A. Srinivasan, and W. Almuhtadi, "QoS Performance Testing of Multimedia Delivery over WiMAX Networks," First International Conference on Data Compression, Communications and Processing (CCP), 2011, vol., no., pp. 271-274.

[12] S. Alshomrani, S. Qamar, S. Jan, I. Khan and I. A. Shah, "QoS of VoIP over WiMAX Access Networks," International Journal of Computer Science and Telecommunications, April 2012, Vol. 3, Issue 4, PP.

[13] T. Wiegand, L. Noblet, and F. Rovati ," Scalable Video Coding for IPTV Services," IEEE Transactions on Broadcasting, JUNE 2009, VOL.55, NO.2, PP.

[14] IEEE Std. 802.16-2004: Part 16: Air interface for fixed broadband wireless access systems [Online]. Available: http://standards.ieee.org/getieee802/802.16.html

[15] M. Chatterjee, S. Sengupta, and S. Ganguly, "Feedback-Based real-time streaming over WiMax," IEEE Wireless Communications Magazine, Feb 2007, vol. 14, no. 1, pp. 64–71.

[16] K. Ahmed, and A. C. Begen, "IPTV and Video Networks in the 2015 Timeframe: The Evolution to medianets," IEEE Communications Magazine, December 2006.

[17] H. Koumaras, C. Skianis, G. Gardikis, and A. Kourtis, "Analysis of H.264 Video Encoded Traffic," Proceedings of the 5th International Network Conference (INC2005), 2005, pp. 441–448.

[18] Intel Co., "H.264 & IPTV over DSL Enabling New Telco Revenue Opportunities," White paper, august 2007. Available at: http://envivio.com/pdf/whitepaper_H264_IPTV_Over_DSL.pdf

[19] K. Ozdemir, R. Jain, A. Moskowitz, K. Ramadas, and M. Vafai, "Triple Play Services including Mobile TV, VoIP, and Internet over Mobile WiMAX Networks," Journal on Optimization, Citeseer, 2009, Vo. , No. , PP. 1-14.

[20] A. Klein, and J. Klaue, "Performance Evaluation Framework for Video Applications in Mobile Networks," International Conference on Advances in Mesh Networks, 2009.

[21] J.R. Ohm, "Multimedia Communication Technology," Springer, USA, 2004.
94

**Jamil M. Hamodi** is currently a Ph.D. candidate in the department of Information Technology at Shri Guru Gobind Singhji Institute of Engineering and Technology (SGGSIET) at Swami Ramanand Teerth Marathwada University, Nanded, India. He received his M.S. degree in Computer networks from the Computer Engineering Department of King Fahd University of Petroleum & Minerals in 2010, Dhahran, Saudi Arabia. His research interests include algorithms, network design and multimedia deployment over IP networks, and 4G wireless (Wimax).

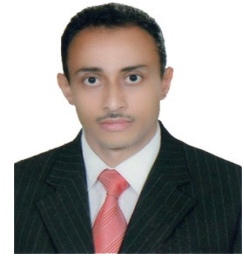

**Dr. Ravindra C Thool** received his B.E and M.E Electronics from SGGS Institute of Engineering and Technology, Nanded (Marathwada University, Aurangabad) in 1986 and 1991 respectively. He obtained his Ph.D. in Electronics and Computer Science from SRTMU, Nanded. Presently he is working as Professor and Head of Department of Information Technology at SGGS Institute of Engineering and Technology, Nanded. His areas of research interest are Computer Vision, Image Processing and Pattern Recognition, Networking and Security. He has published more than 50 papers in National and International Journals and Conferences. Two students have completed their Ph.D. and 10+ students are working for their Ph.D. under his guidance. He is the member of IEEE, ISTE, ASABE (USA) and also a member of University Co-ordination committee of NVidia. Also, working as Expert on AICTE committee and NBA.

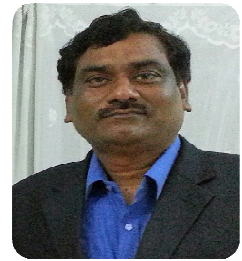